\documentclass[preprint2]{aastex}

\shorttitle{UCAC3 atrometric reductions}
\shortauthors{Finch}

\begin{document}

\title{UCAC3: Astrometric Reductions}

\author{Charlie T. Finch, Norbert Zacharias, Gary L. Wycoff}

\email{finch@usno.navy.mil}

\affil{U.S. Naval Observatory, Washington DC 20392--5420}


\begin{abstract}

Presented here are the details of the astrometric reductions from the
$x,y$ data to mean Right Ascension (RA), Declination (Dec) coordinates
of the third U.S. Naval Observatory (USNO) CCD Astrograph Catalog
(UCAC3).  For these new reductions we used over 216,000 CCD exposures.
The Two-Micron All-Sky Survey (2MASS) data are extensively used to
probe for coordinate and coma-like systematic errors in UCAC data
mainly caused by the poor charge transfer efficiency (CTE) of the 4K
CCD.  Errors up to about 200 mas have been corrected using complex
look-up tables handling multiple dependencies derived from the
residuals.  Similarly, field distortions and sub-pixel phase errors
have also been evaluated using the residuals with respect to 2MASS.
The overall magnitude equation is derived from UCAC calibration field
observations alone, independent of external catalogs.  Systematic
errors of positions at UCAC observing epoch as presented in UCAC3 are
better corrected than in the previous catalogs for most stars.  The
Tycho-2 catalog is used to obtain final positions on the International
Celestial Reference Frame (ICRF).  Residuals of the Tycho-2 reference
stars show a small magnitude equation (depending on declination zone)
that might be inherent in the Tycho-2 catalog.

\end{abstract}


\section{INTRODUCTION}

The U.S.~Naval Observatory (USNO) CCD Astrograph Catalog (UCAC)
project began observations in the Southern Hemisphere at Cerro Tololo
Interamerican Observatory (CTIO) in January 1998.  In October 2000 the
first U.S. Naval Observatory CCD Astrograph catalog (UCAC1)
\citep{2000AJ....120.2131Z} was published covering about 80\% of the
southern sky with positions and preliminary proper motions for over 27
million stars.  The Astrograph was moved to the Naval Observatory
Flagstaff Station (NOFS) in October 2001 to complete the Northern
Hemisphere observing after 2/3 of the sky were completed from CTIO.
The second USNO CCD Astrograph catalog (UCAC2)
\citep{2004AJ....127.3043Z} was released in July 2003 with the same
level of completeness as in UCAC1, but with improved reduction
techniques, early epoch plates for improved proper motions and
extended sky coverage.  All sky coverage for UCAC observations were
completed in October 2004.

The third USNO CCD Astrograph Catalog (UCAC3) \citep{u3r}
is the first all-sky data release in the UCAC series,
containing about 100 million entries with a slightly
fainter limiting magnitude than in previous versions.  The magnitude
range for UCAC3 is roughly 8.0 to 16.3 mag in the UCAC bandpass
(579-642 nm, hereafter UCAC magnitude). 
A detailed introduction into UCAC3 with comparisons to other catalogs
and warnings for the users are given in \citep{u3r}.
Any user is also urged to read the extensive ``readme'' file
provided with the DVD or on-line release.

The UCAC3 is based on a complete re-reduction of the pixel data aiming
at more completeness with the inclusion of double star fitting,
problem case investigations and the slightly deeper limiting magnitude
\citep{u3x}.  The final positions are based on the Tycho-2
\citep{2000A&A...355L..27H} reference frame as in UCAC2.  However,
Two-Micron All Sky Survey (2MASS) \citep{2006AJ....131.1163S}
residuals are used to probe for systematic errors in astrometric
reductions as will be explained below.

In UCAC1 and UCAC2 it was shown that the 4K CCD in the astrograph
camera has a relatively poor charge transfer efficiency (CTE),
leading to coma-like systematic errors in the uncorrected stellar positions.
The effect is seen mainly in the $x$-axis (right ascension), which is the 
direction of the fast readout of charge, while the $y$-axis (declination)
shows a much smaller effect.  In UCAC1 a simple empirical approach was used
to correct for this effect with the basic assumption that the effect was
linear along the $x$-axis and no assumption for a dependency on
magnitude.  For UCAC2 the empirical approach was extended with a more
complex model as a function of $x,y$, and instrumental magnitude
derived from flip observations of calibration fields.  Flip observations
are obtained by observing the same field with the telescope on one side
of the pier (east or west) then repeat with the telescope on the other side.
Thus two images of the same area in the sky are obtained which are
rotated by $180^{\circ}$ with respect to each other.

In both previous UCAC catalogs sub-pixel phase and field distortion
errors have also been investigated.  For UCAC1 the pixel phase was
modeled with an empirical function showing an amplitude on the order
of 12 mas resembling a sine-function. For UCAC2 the pixel phase errors
were investigated further and found to also be a function of the full
width at half maximum (FWHM) of the stellar image profiles.  
The field distortion pattern was modeled in both previous reductions 
by binning reference star residuals from individual frames used in 
the reductions.

The new pixel data reductions up to $x,y$ data are described in
\citep{u3x}.  This paper describes the reductions following the $x,y$
data up to the CCD-based mean RA, Dec positions.  Early epoch data and
procedures to derive proper motions are presented in the UCAC3 release
paper \citep{u3r}, while details about an important part of the early
epoch data, the Southern Proper Motion (SPM) data will be given
elsewhere \citep{spm}.

\section {INPUT DATA}

For the astrometric reductions of UCAC3 as described in this paper,
two sets of input data are needed: the $x,y$ data from selected CCD
observations, and reference stars.  
Here two different reference star catalogs are used for different purposes.

\subsection {CCD Frame Selection}

Out of the about 278,000 UCAC frames ever taken, all applicable survey
frames, calibration field, and minor planet exposures are selected for
the reductions presented here.  Poor quality frames are excluded.
Quality criteria include limiting magnitude, internal fit precision
of high S/N stellar images, and mean image elongation.  About 15\% of the
observations are qualified as ``poor", see also \citep{2000AJ....120.2131Z}.
All frames which pass those quality criteria are included. Then
the all-sky completeness from the selected data are examined and
frames which almost meet the quality criteria are included as needed 
to provide a complete all-sky coverage.

The final UCAC3 catalog contains mainly the survey frame data.
The minor planet observations will be published separately,
while the calibration field observations are only used in the first
reduction steps to derive corrections to systematic errors.
A summary of the CCD observations is given in Table~\ref{framest1}.
The frames taken along the path of Pluto are included here from
a collaboration with L.~Young (SWRI) for occultation predictions.

A total of about 50 fields in the sky observed at about 10 to 30
different epochs with multiple exposures each time were used as
astrometric calibrations.  These typically low galactic latitude
fields around ICRF targets, equatorial calibration fields, and open
clusters were observed about 10 to 30 times during the project and
with the telescope flipped between orientations east and west of the
pier to provide a reversal of RA, Dec orientation in $x,y$ space.
Data from these fields are utilized to derive certain systematic error
corrections (see below).  However, these frames are not included in
the final UCAC3 reductions based on the Tycho-2 reference catalog.

Frames around extragalactic link sources taken with the astrograph at 
times of deep CCD observations (mostly with the KPNO and CTIO 0.9 m 
telescopes) are not included here either. 
These data are still under investigation and a separate paper is
in preparation regarding the optical link to ICRF quasars.

\subsection {$x,y$ Data}

For all astrometric reductions described below only the $x,y$ results
from pixel data profile fit model five (symmetric Lorentz profile with 
pre-set $\alpha, \beta$ shape parameters) are used.  
This fit model has five free parameters per single star image 
(background, amplitude, center $x,y$, width of profile),
similar to the more familiar two-dimensional Gauss model.
However, the Lorentz profile matches the observed point-spread function
(PSF) of our data significantly better than the Gaussian model.
For more details and explicit PSF model functions see the UCAC3 pixel 
reduction paper \citep{u3x}.

Double star fits of blended images are based on the same Lorentz image
profile model, however three more free parameters are used for the two
center coordinates and the amplitude, respectively, of the secondary
component.  Both the primary and secondary component are fit
simultaneously.  A single width of the profiles for both components of
a double star and a single background level parameter is used in this
fit.

The $x,y$ data files also contain internal errors on the center
coordinates derived from the least-square fits, two instrumental
magnitudes based on the profile fit model and a real aperture photometry,
respectively, and several auxiliary flags from the raw pixel reduction step.
For more information about the raw data reductions see the UCAC3 pixel 
reduction paper \citep{u3x}.

\subsection {Reference Stars}

The 2MASS catalog does not provide proper motions.  However, the UCAC
and 2MASS observations were made almost at the same epoch and the
2MASS positions are of high quality with random errors of about 70 mas
per coordinate for well exposed stars and small systematic errors 
\citep{jd16}.  
Due to the deep limiting magnitude and high density the 2MASS catalog 
is an excellent tool to probe UCAC data for systematic errors on a
statistical basis, allowing to stack up many residuals as a function of 
a large number of parameters, as will be described below.
Table~\ref{framest2} lists the number of available observations of
reference stars, each providing a residual along RA ($x$) and Dec ($y$).

For this purpose a subset of the 2MASS point source catalog is
constructed.  Stars are selected from the Naval Observatory Merged
Astrometric Dataset (NOMAD) using the 2MASS identifier and imposing 
a limit of V or R $\le$ 16.5 to select 116,247,341 2MASS stars.
For these, the original 2MASS positions are retrieved and
matched with UCAC observations on a frame-by-frame basis.

Proper motions of this reference star catalog are taken out of
the NOMAD catalog.  Even if these are not very accurate for faint
stars, they bridge the few years of epoch difference to UCAC data
to avoid large-scale biases from galactic dynamics.
For most stars in the southern hemisphere, the epoch
difference between 2MASS and UCAC is $\sim \pm$ 1 yr, while it
gradually increases for stars at higher declinations.  
Near the north celestial pole the epoch difference is about four years.


The 2MASS data are used only to derive most of the systematic error
corrections of the UCAC $x,y$ data.  As with UCAC2, the Tycho-2
catalog is used as the only source for reference stars in a final
astrometric reduction to obtain UCAC3 positions on the International
Celestial Reference Frame (ICRF).  Only stars from the Tycho-2 catalog
indicated as having good astrometry have been used for this reduction.
Tycho-2 proper motions have been used to propagate the positions to
the epoch of the individual UCAC $x,y$ observations.

Both reference star catalogs are sorted by declination and
stored in binary direct access files.  For each catalog a match to the
$x,y$ data are performed to produce cross-reference output files per CCD 
frame containing the record numbers from the $x,y$ direct access data and
the reference star catalog, as well as the magnitude of the star,
B$-$V color, and astrometry flag (for Tycho-2).  This scheme allows a
fast runtime for the many passes through all the data to perform the
astrometric reductions without the need for a match of reference stars
each time.

\section {ASTROMETRIC REDUCTIONS}

For the astrometric reductions in UCAC3 extensive systematic error
investigations are performed.  The largest systematic error is caused
by the poor charge transfer efficiency (CTE) of the detector, followed
by geometric field angle distortions.  An iterative approach was
adopted utilizing 2MASS residuals to determine each of these effects
in turn as described below.  The idea here is to use an empirical
approach as done in the past, but investigate further dependencies to
better model the effects seen in the 4K CCD pixel data.  Preliminary
results were presented earlier \citep{2009AAS...21347305F}.  A pure
magnitude equation is derived from internal calibration observations
only.  Finally the position shifts as function of the sub-pixel
location of an image in the focal plane (the sub-pixel phase error) is
derived, however, the $x,y$ data used for that are the original
measures before applying the other corrections.

Systematic position offsets as a function of color and differential
color refraction due to Earth's atmosphere are small (about 5 mas)
due to the narrow UCAC spectral bandpass.
No such corrections are applied for the UCAC3 catalog.

\subsection {Preliminary Field Distortion Pattern}

In the first step of the astrometric reductions, the CTE caused
systematic errors (coma-like) are approximated by a linear model as a
function of magnitude and $x$ pixel coordinate, similar to the CTE
modeling of UCAC1.  This takes out about 80\% of the CTE effect.
The pre-corrected $x,y$ data are then used in a
preliminary astrometric reduction with 2MASS reference stars to produce
an approximate field distortion pattern (FDP), i.e.~systematic errors
purely based on the $x,y$ location of a stellar image on the area of
the detector.
Although the coma-like corrections for the CTE effect should not
bias the purely geometric FDP corrections, the scatter in the data
is largely reduced by correcting for most of the CTE effect first.
This leads to a more accurate FDP as would be generated without 
applying any CTE corrections.

Maximal systematic errors in the FDP are about 24 mas, with typical
corrections in the 5 to 10 mas range.  This preliminary FDP is then
applied to the otherwise uncorrected $x,y$ data to analyze the CTE
effect from scratch in the following step.
The same reasoning applies here; with good FDP correction in hand
the scatter of the residuals is reduced to accurately probe the
systematic errors induced by the poor CTE.

\subsection {CTE Effect}

When the CCD reads out an image, the electrons are transferred from
pixel$-$to$-$pixel until the charge reaches the output register.  The
low CTE of the 4K CCD causes charge to be left behind as this transfer
occurs, leading to slightly asymmetric images.
The amount of asymmetry and the derived $x,y$ center of stellar images
with respect to the unaffected position depend on the $x$ pixel location,
the brightness of the star, the length of the exposure, and other factors.

\begin{figure}
\epsscale{1.00}  
\plotone{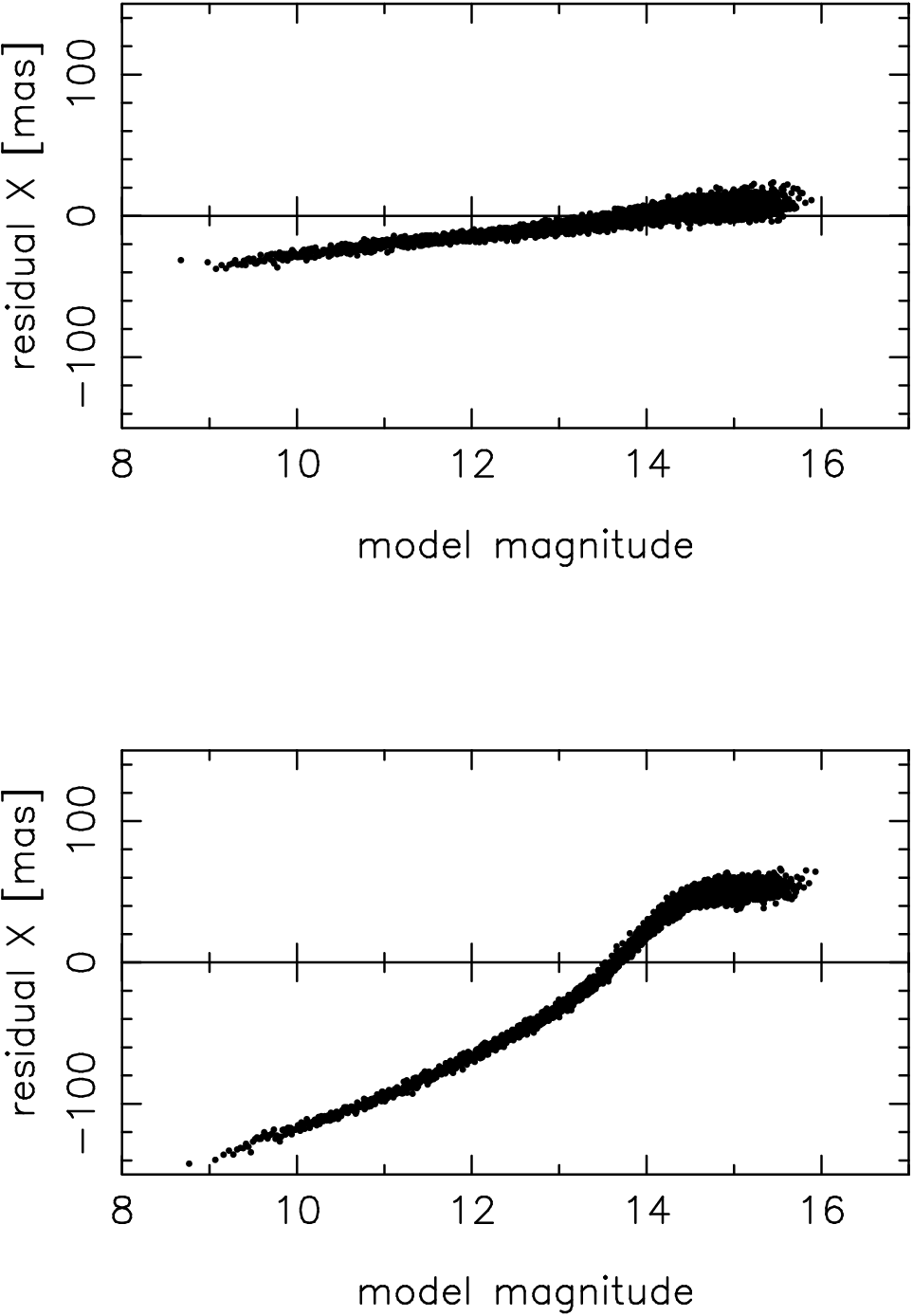} 
\caption{CTIO west residuals in $x$ with respect to 2MASS
reference stars as a function of UCAC model magnitude using frames 
taken at 20 second exposures (short).  The top plot shows the residuals
for low $x$ near pixel 1 and the bottom plot for high $x$ near 
pixel 4094.  Each dot represents the mean over 1000 residuals.}\label{CTEf1}
\end{figure}

As found here and in previous UCAC reductions the CTE effect is by far
the most substantial systematic error seen in the raw UCAC $x,y$ data
amounting up to about 200 mas.  The effect is predominantly seen in
the $x$-axis along right ascension and increases from no effect near
$x = 0$ pixel to a maximum near $x = 4094$ pixel, as evident in
Figure~\ref{CTEf1}.  A similar effect is also seen in the $y$-axis,
but to a lesser degree because of a slower charge transfer in the
direction of declination.

For the reductions, the UCAC frames are separated into individual data
sets depending on observation site (CTIO or NOFS) and telescope
orientation (east or west).  Over 216,000 frames are used for
the reductions, see Table~\ref{framest1}.  Most frames at CTIO were
observed with the telescope west of the pier, while at NOFS the
regular observing was performed east of the pier.

Because the instrument had to be disassembled for the move from
CTIO to NOFS the camera alignment and other instrument properties 
are slightly different at both sites.
This explains the slightly different patterns for systematic
error corrections found for the two sites.
However, the 2MASS reference star catalog is dense enough to
provide a sufficient number of residuals for statistically significant
results even on relatively small sub-sets of the UCAC data.

\begin{figure}
\epsscale{1.00}
\plotone{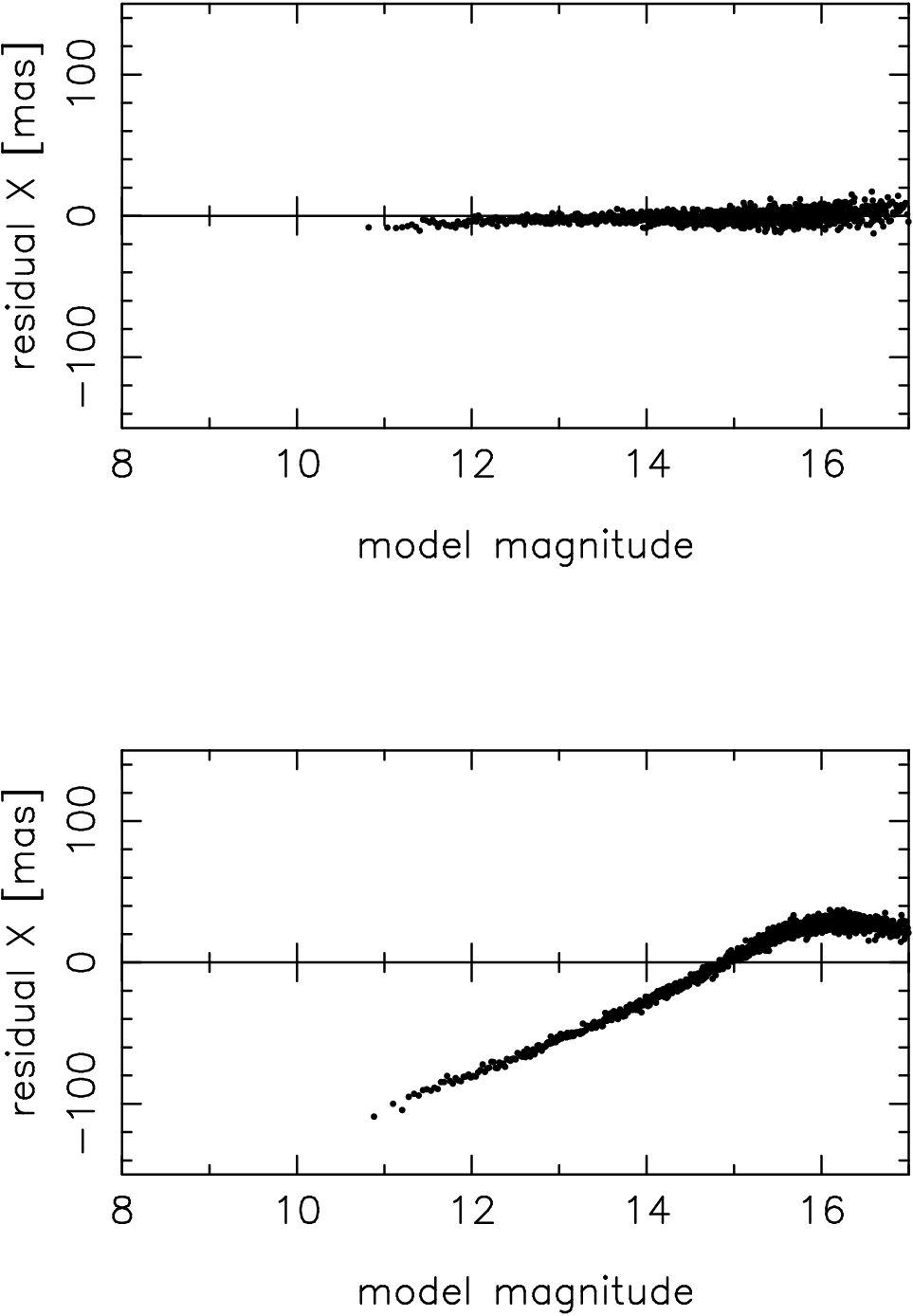}
\caption{CTIO west residuals in $x$ with respect to 2MASS reference
stars as a function of UCAC model magnitude using frames taken at 150
second exposures (long).  The top plot shows the residuals for low $x$
near pixel 1 and the bottom plot for high $x$ near pixel 4094.  Each
dot represents the mean over 1000 residuals.}\label{CTEf2}
\end{figure}

For UCAC3 we found that the CTE systematics show a dependence on FWHM,
brightness of the star, exposure time, location on the chip ($x,y$),
camera orientation (east, west) and observing site (CTIO, NOFS). 
For example as Figures~\ref{CTEf1} and \ref{CTEf2} show the CTE caused
different systematic position offsets for the $x$ coordinate as a function
of magnitude for 20 and 150 sec exposures, respectively.

The residuals with respect to the 2MASS reference stars
are split into two major data sets for investigating the CTE effect,
CTIO west and NOFS east.  A plotting program is then used to display
and determine empirical corrections to create a complex look-up
table.  The table is split up into four different FWHM bins keeping a
roughly equal number of frames per bin, with half step magnitude bins 
ranging from 8 to 16 magnitude for all standard exposure times of
20$-$200 seconds duration, and 5 bins along the $x,y$ axes.  
Each data set is evaluated and look-up tables are created to correct 
for the residuals, (see sample, Table~\ref{CTEt1}).  

For each of the main data sets, CTIO west and NOFS east, tables are 
created separately for the $x$ and $y$ axes.  After
testing we found that these tables could also be used for the data of
the other configurations, (CTIO east and NOFS west) corresponding to
the observation site.  After applying corrections and re-running the
residuals the correction tables are continuously updated until the
residuals flattened out.  The largest correction from the residuals
for CTIO is 204 mas in the $x$-axis and $-$32 mas in the $y$-axis.
For NOFS the largest correction from the residuals is 216 mas in the
$x$-axis and $-$67 mas in the $y$-axis.

\subsection {Pure Magnitude Equation}

The 2MASS positions are uncorrelated to the $x,y$ pixel coordinates 
of the UCAC observations.  Thus any systematic errors seen in the
2MASS$-$UCAC residuals as a function of UCAC $x,y$ and also as a
function of magnitude times these coordinates (coma terms) are
inherent in the UCAC data and can be corrected with the above
procedure.

\begin{figure}
\plotone{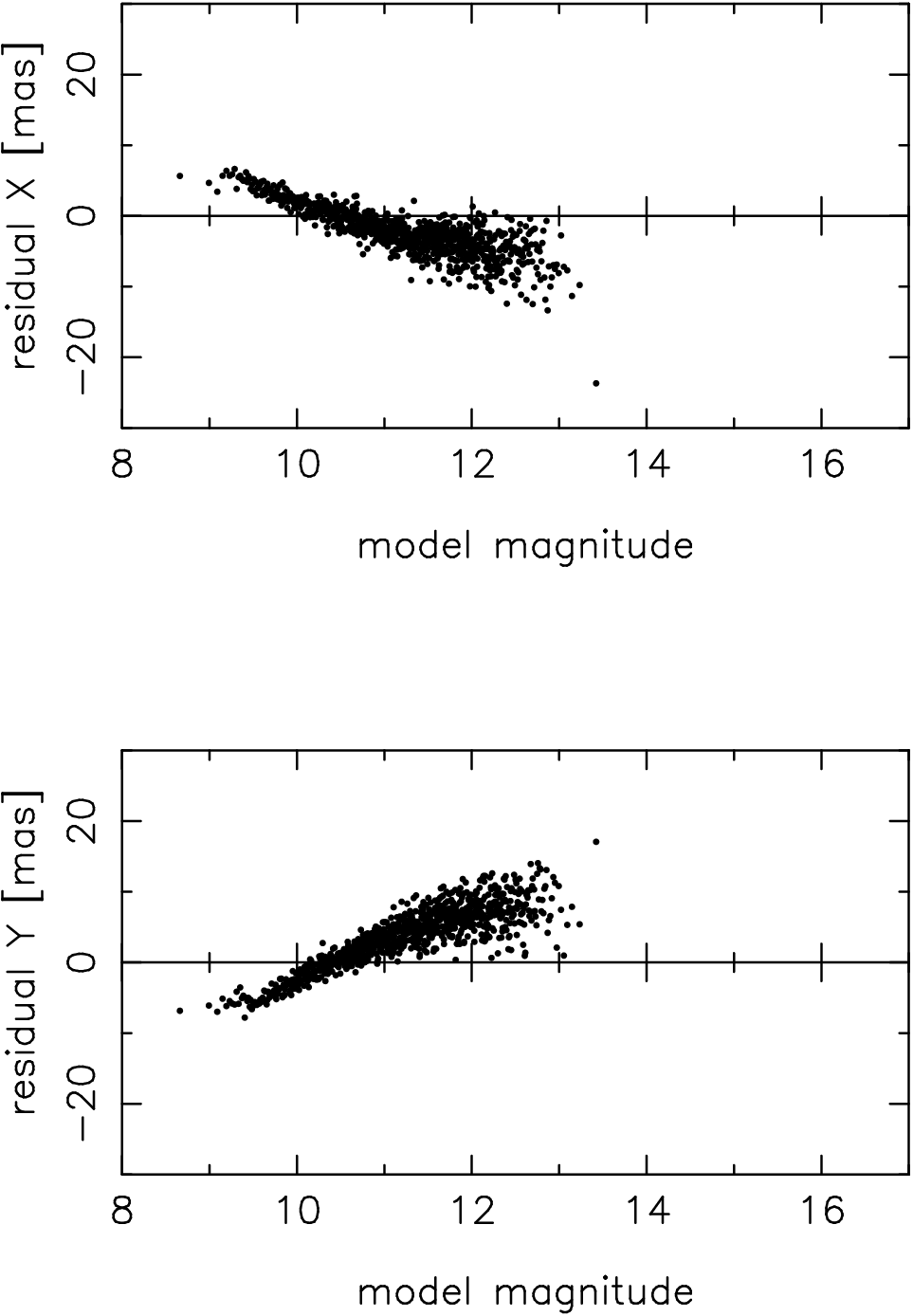}
\caption{Residuals in $x$ (top) and $y$ (bottom) for
CTIO west with respect to Tycho-2 reference stars as a function of
UCAC model magnitude.  Each dot represents the mean over 3000
residuals. }\label{TYCf1}
\end{figure}

However, this is not the case for a pure magnitude dependent
systematic error, which could originate in either or both catalogs.
With the systematic error corrections applied to the UCAC $x,y$ data
as described above any possible pure magnitude equation from 2MASS
was transferred into the UCAC positions.  Different catalogs
have typically different magnitude equations.  As an example
Figure~\ref{TYCf1} shows the residuals as function of magnitude from a
reduction of UCAC data with Tycho-2 reference stars, after applying
all systematic error corrections based on the 2MASS reductions.  
This clearly shows a difference in magnitude equation between 2MASS
(= UCAC system at this point) and
Tycho-2 without knowing what the error free positions might be.

The flip observations of 
calibration fields are used to determine the overall, pure magnitude
dependent systematic errors in the UCAC data independent of any
external reference star catalog.  With these observations the same
field in the sky has been observed with sets of frames rotated by
180$^{\circ}$ with respect to each other.  Any $x,y$ coordinate offset
as function of magnitude shows up in the residuals of a transformation
of east versus west frames $x,y$ data.  We derived overall magnitude 
equation slope terms for long and short exposure frames and both
sites separately.  These are then applied globally in the final 
astrometric reductions.

After applying all corrections a small magnitude term of a few 
mas/mag is still seen in the residuals of the Tycho-2 reductions 
of UCAC data as shown in the UCAC3 release paper \citep{u3r}.  
This indicates such a magnitude
equation is present in the Tycho-2 catalog itself, which is found to
vary as a function of declination zone as one would expect.

It would be preferable to derive all systematic error corrections from
internal calibration observations.  However, the flip observations
alone do not allow us to do so because of the degeneracy between a
pure magnitude equation and coma-like terms.  Only after correcting
the UCAC $x,y$ data for coordinate dependent (including coma) terms do
the flip observations allow for a unique solution of the magnitude
equation.  Of course the assumption here is a constant magnitude
equation, not changing from exposure to exposure.  This assumption can
not easily be made for photographic astrometry, which is affected by a
highly non-linear detector, but should hold better for CCD data.

\subsection {Field Distortions}

Field distortion patterns (FDPs) are derived for UCAC3 by binning
thousands of reference star residuals of individual CCD frames using
the same procedure as in the previous UCAC reductions.  These
reductions are performed on the CTE corrected data but without
applying the preliminary FDP used before.

From deriving FDPs of various subsets of the data we found that the
FDP is almost constant except for a small difference depending on
observing site (CTIO or NOFS).  Residuals are created using all good
survey frames with respect to 2MASS reference stars split into CTIO
west and NOFS east data sets.  The data using opposite camera
orientations than used for most of the observations (i.e.~CTIO east
and NOFS west) did not have significantly different field distortions
so only two correction tables are used for the final reductions.  FDP
corrections for frames taken with the camera orientation not used
frequently are applied by rotating the FDP of the data with the large
amount of observations at that site.

\begin{figure}
\includegraphics[angle=270,scale=0.5]{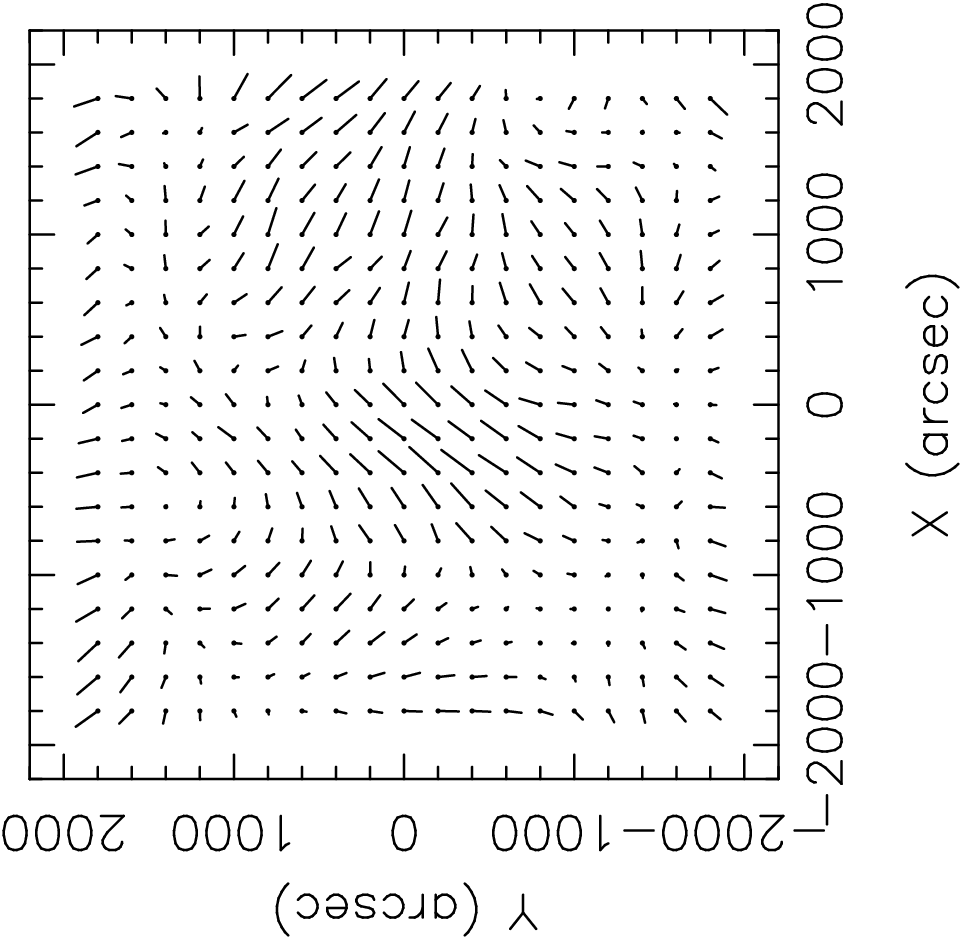}
\includegraphics[angle=270,scale=0.5]{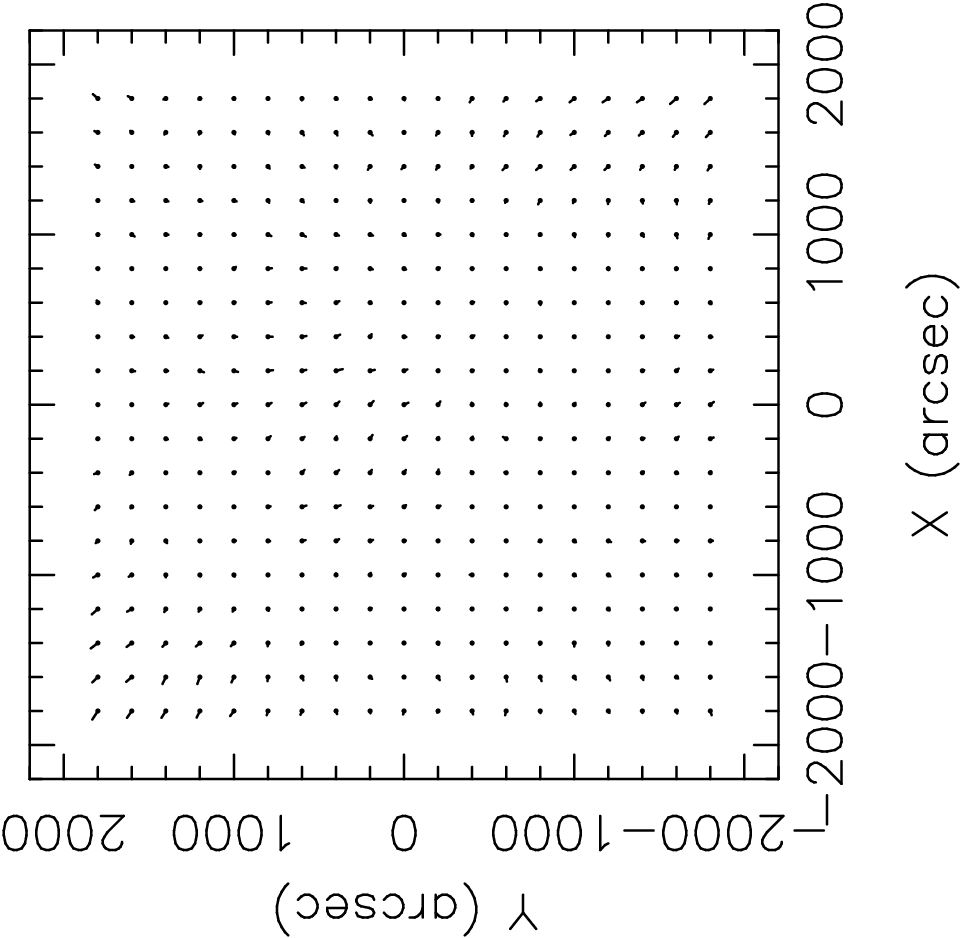}
\caption{Field distortion pattern (FDP) plot of 2MASS stacked 
residuals for CTIO west (top) and
the difference between vectors of CTIO west and NOFS east (bottom).
The scaling of the vectors is 10,000 which makes the largest
corrections for CTIO west (top) 23 mas and the largest difference
vector (bottom) 6 mas.}\label{FDPf1}
\end{figure}

Figure~\ref{FDPf1} (top) shows the FDP for CTIO west with vectors up
to 23 mas in length.  The FDP for NOFS east is slightly different with
vectors up to 24 mas at some bins.  In Figure~\ref{FDPf1}
(bottom), we show the difference (CTIO west$-$NOFS east) of the two  
data sets.  Although these differences are small (below 6 mas) they 
are systematic and well determined.  Therefore we choose to use a 
separate FDP map to correct the data of each site.

\subsection {Subpixel Phase Errors}

\begin{figure}
\epsscale{1.00}
\plotone{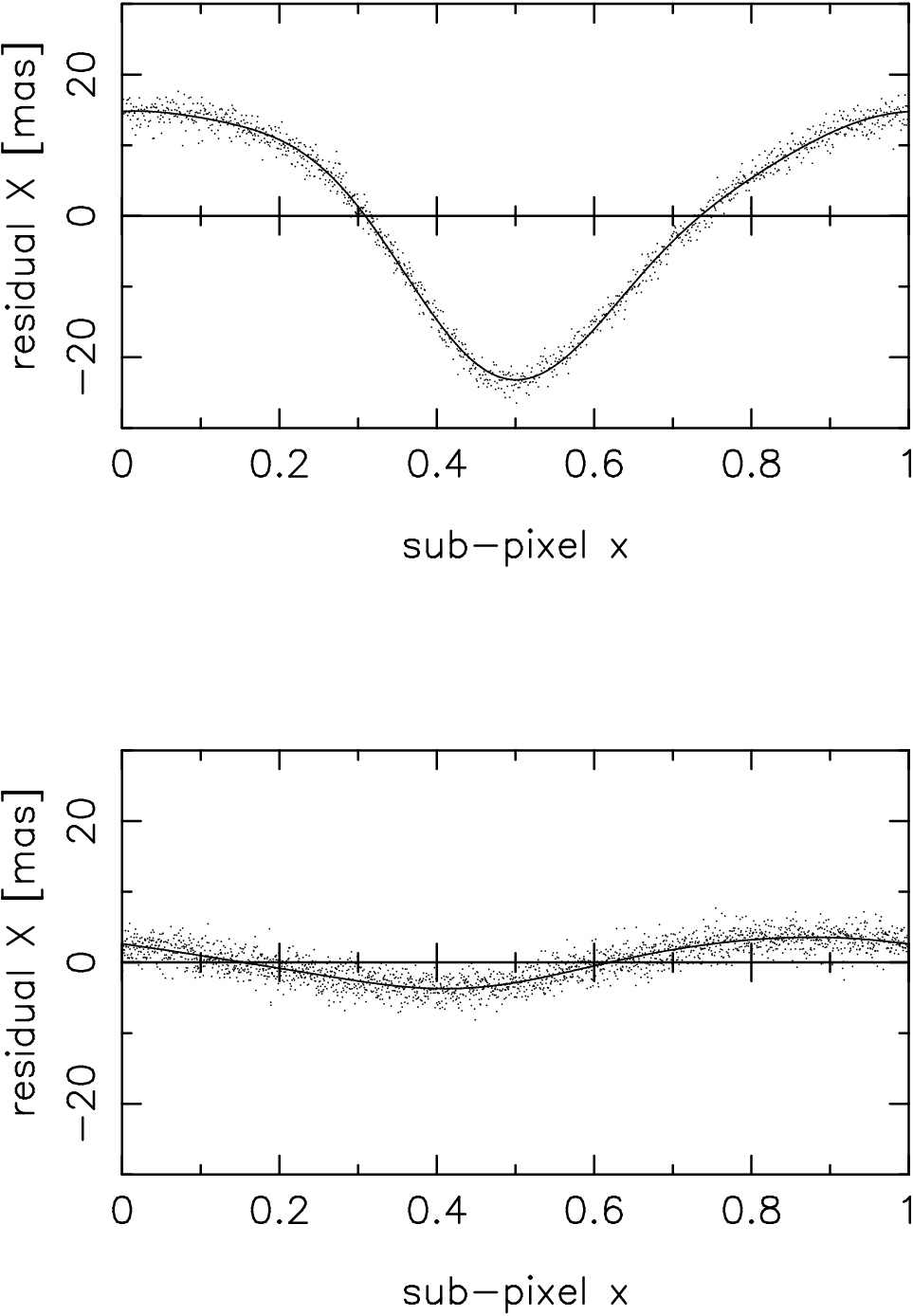}
\caption{CTIO west residuals in $x$ with respect to 2MASS
reference stars as a function of sub-pixel phase. 
The top and bottom plot show residuals with
an average FWHM of 1.54 and 2.11 pixel respectively.  
Each dot represents the mean over 5000 residuals.
The fitted curve is from a least-squares adjustment using a model
with a total of six Fourier terms.}\label{SUBPf1}
\end{figure}

After the above mentioned systematic errors have been corrected the
2MASS reference star catalog is used again to generate residuals from
all applicable UCAC survey frames.  Residuals are analyzed as a
function of the original $x$ and $y$ pixel coordinate fraction
(sub-pixel phase) before other corrections are applied.  Various
sub-sets of the data are looked at.  Systematic errors are found to be
a function of the FWHM of the image profiles.  Figure~\ref{SUBPf1}
gives some examples.  The results are found to be independent of
exposure time, as expected.  However, a slight difference between the
CTIO and NOFS data is found.

The amplitude of the sub-pixel phase dependent systematic errors in the
star positions is shown in Figure~\ref{AMPf1},
here for the corrections to the $x$ coordinate;
those for the $y$ coordinate are somewhat smaller.
All sub-pixel phase systematic corrections are smaller than what was
found in UCAC2.  
This is a consequence of using an image profile model in
UCAC3 (Lorentz profile) which better fits the true PSF than was the
case for UCAC2 (Gauss model).

However, the function of the sub-pixel phase systematic errors are
more complex in the UCAC3 than UCAC2 data, where a simple sine and
cosine term were sufficient.  For the UCAC3 data we had to expand to
three sine and three cosine terms in order to fit the sub-pixel phase
errors sufficiently well (Figure~\ref{SUBPf1}).  These six parameters
are determined separately for 12 sets of data binned by FWHM (from 1.5
to 3.0 pixels), and split by NOFS and CTIO data.  Calibration tables
are then generated for equal steps along FWHM by interpolation and
$x,y$ corrections applied, separately for each coordinate, based on
these tables.

\section {MEAN POSITIONS} 

Positions of all detected objects are obtained frame-by-frame from a
final astrometric reduction with the Tycho-2 reference star catalog
and correcting raw $x,y$ data first for the sub-pixel phase errors,
then for systematic errors as a function of $x,y$ (FDP), then for
mixed terms of coordinate with magnitude (CTE effect), and finally for
a pure magnitude equation, as explained above.  Apparent places and
refraction are corrected rigorously using the Software for Analyzing
Astrometric CCD (SAAC) code \citep{saac}, which also utilizes the
Naval Observatory Vector Astrometry Subroutines (NOVAS) code
\citep{novas}\footnote{$http://aa.usno.navy.mil/software/novas/novas\_info.php$}.
The thus obtained positions are on the ICRF at individual epoch of
each CCD frame (between 1998 and 2004) and are output to FPOS (final
position) files.

Previously identified and flagged observations of minor planets and
high proper motion stars are output to separate files.  All other
individual positions are output by declination zones and then sorted
by declination.  Weighted mean positions are calculated from the
individual images of each star, generating a running star number, 
MPOS (mean position file) on the fly.  Over 139 million objects are
identified at this step.

All MPOS entries are then matched with early epoch star catalogs and
another, more comprehensive 2MASS extract containing about 338 million
objects.  These 2MASS stars are selected directly from the 2MASS point
source catalog without going through NOMAD.  The R magnitude was
estimated based on the 2MASS near-infrared J$-$K color and stars with
R $\le$ 17.0 or J $\le$ 15.5 are selected.  The unique identifier for
stars matched across catalogs is the MPOS star number.  Individual
early epoch positions are output together with MPOS entries (CCD epoch
observations) and sorted by MPOS number.  Weighted proper motions and
mean positions are then calculated to obtain the UCAC3 release catalog
data \citep{u3r}.  Objects which did not have either a reasonable
proper motion determined or could not be matched with 2MASS are
dropped at this point.  Only these compiled catalog mean positions and
proper motions are published in UCAC3, not the MPOS or FPOS data,
which likely will be made available for the final UCAC4 release after
further updates.

\section{COMPARISON WITH HIPPARCOS}

A total of 1510 Hipparcos stars in the 8 to 12 magnitude range are
randomly selected (about 300 all sky per magnitude interval) and
flagged in the UCAC data.  These stars are not used as reference
stars in test reductions using Tycho-2 reference stars.  Thus the
obtained positions are field star positions from UCAC observations
independent of the Hipparcos and Tycho catalog positions.
Individual UCAC observed positions are then compared to the original 
\citep{hiporig} and new Hipparcos reductions \citep{2007ASSL..350.....V} 
at the epoch of UCAC observations, using Hipparcos mean positions, 
proper motions and parallaxes.

\begin{figure}
\epsscale{1.00}
\plotone{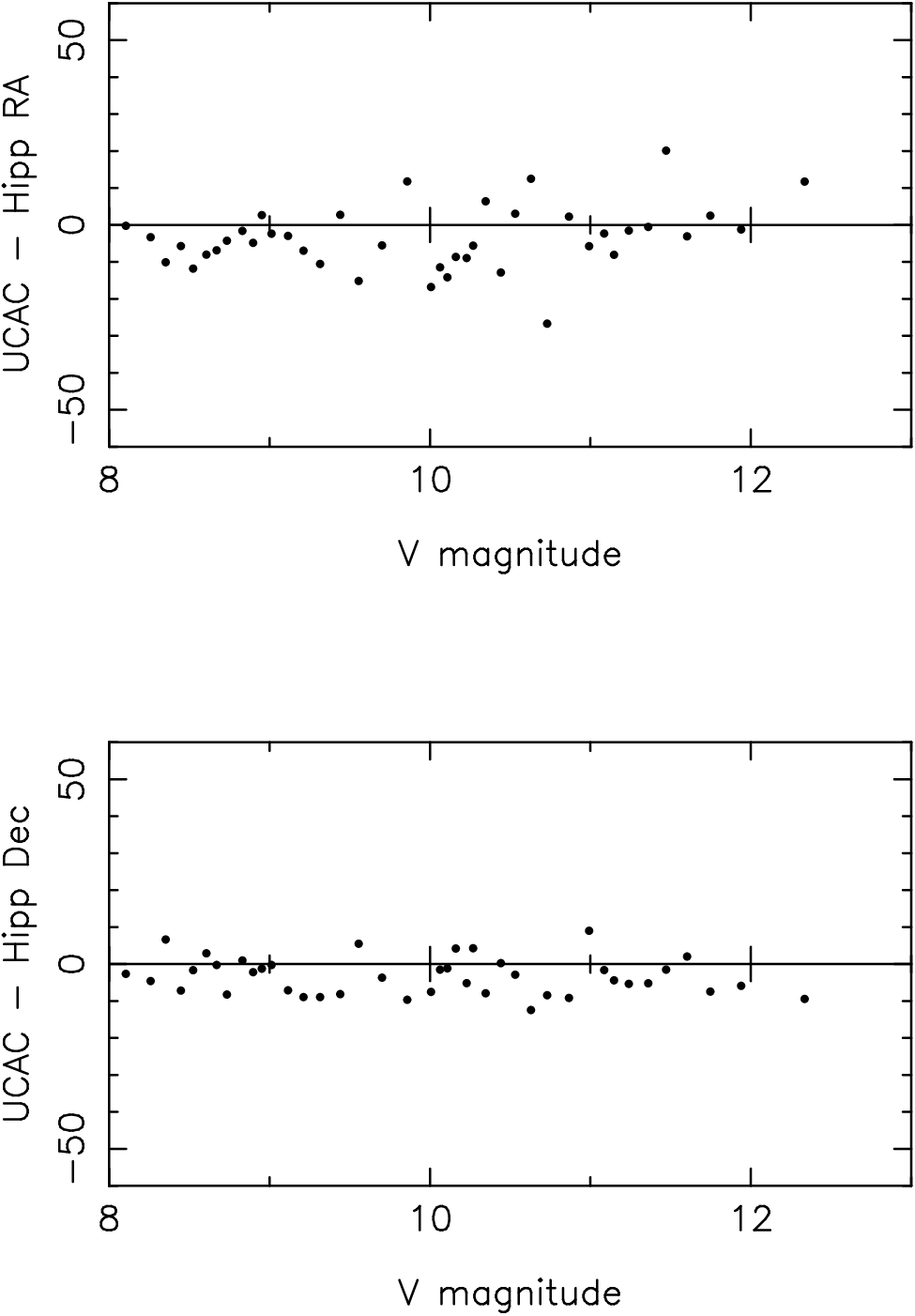}
\caption{Position differences UCAC$-$Hipparcos (new reductions)
as a function of V magnitude of a random sample of Hipparcos stars 
reduced as field stars in UCAC processing.  Each dot is the mean of 
100 individual UCAC observations.}\label{UHipmag}
\end{figure}

\begin{figure}
\epsscale{1.00}
\plotone{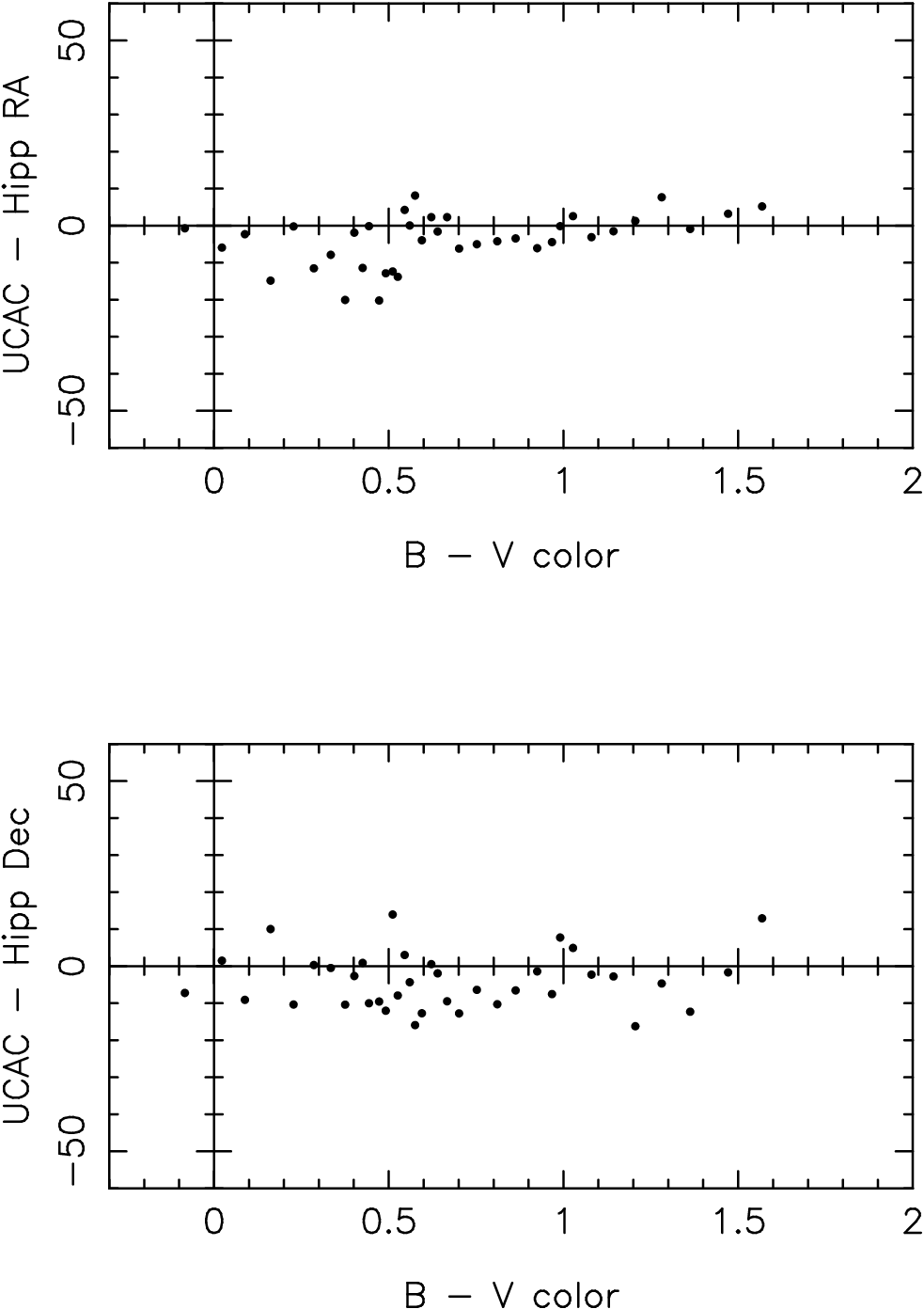}
\caption{Position differences UCAC$-$Hipparcos (new reductions)
as a function of B$-$V color of a random sample of Hipparcos stars 
reduced as field stars in UCAC processing.  Each dot is the mean of 
100 individual UCAC observations.}\label{UHipcol}
\end{figure}

After excluding outliers ($\ge$ 200 mas position difference in either
coordinate), RMS values over observations in each bin are calculated
(Table~\ref{TYCt1}).  Similarly sorting all
observations by magnitude or color, respectively and binning over 100
observations lead to the plots shown in Figure~\ref{UHipmag} and
\ref{UHipcol}.

The expected position errors from UCAC3 and Hipparcos data at the
epoch of our UCAC observations are also presented in Table~\ref{TYCt1}
together with the expected RMS of the combined error and the ratio of
expected to observed scatter, separately for each coordinate.  In all
cases the observed errors (from the scatter of the UCAC3$-$Hipparcos
position differences) is slightly smaller than the expected errors as
calculated from the combined formal errors for the same observations,
thus at least some of these are overestimated (see also discussion
section below).

UCAC position differences of those sampled Hipparcos stars
do not show systematic errors as a function of magnitude or
color exceeding about 10 mas over the range sampled.  
Plots with respect to the original or new Hipparcos catalog 
are almost identical.

However, 23 Hipparcos stars (1.5\% of this sample) show very large
differences (between 300 and 600 mas in either coordinate) when
comparing the new reduction Hipparcos positions with the UCAC3
positions at UCAC epoch.  A similar number of stars is found
when comparing with the original Hipparcos Catalogue; however, 
for not exactly the same stars.  All possible combinations of 
inconsistencies between the two Hipparcos solutions and UCAC data 
are found, with two of the three positions or all three separated by
several standard errors of their internal errors.

\section {DISCUSSION}

\begin{figure}
\epsscale{1.00}
\includegraphics[angle=-90,scale=.40]{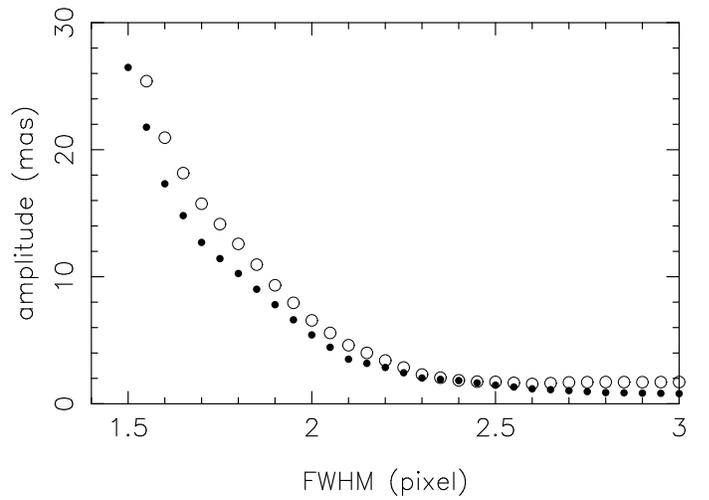}
\caption{Amplitude of the sub-pixel phase dependent positional
correction as a function of image profile width (FWHM) for the CCD
astrograph frames taken at CTIO (filled) and NOFS (open circles),
respectively.}\label{AMPf1}
\end{figure}

The use of an image profile model better matching the actual PSF than
a Gaussian model is essential for the astrometric reductions of
blended images.  A better matching model also does reduce the
amplitude of the sub-pixel phase error and is advisable to be used when
no such corrections are being applied to the data.  In particular, a
comparison of Figure~\ref{AMPf1} with a similar figure of the 
UCAC2 paper show that with a Gaussian model and 2.0
pixel/FWHM sampling the pixel phase error has an amplitude of 11 mas,
while with the image profile model 5 as used in UCAC3 this amplitude is
only about 6 mas.  

The use of such a PSF profile model (still with the same
number of fit parameters per star as the traditional Gaussian model)
allows to neglect positional errors as function of sub-pixel phase 
completely for a sampling of about 2.5 pixel/FWHM or larger without 
the need to investigate possible other dependencies of this systematic
error as a function of other things.
However, for single stars and with calibration data in hand to correct
for the position offsets caused by a sub-pixel phase dependency, the
use of a more sophisticated image profile model than a Gaussian might
not have an apparent advantage for astrometric reductions.

The slight difference in amplitude of the sub-pixel phase corrections
between CTIO and NOFS data is surprising.  It could be caused by a
slightly different, observed PSF between the two sites, even for the
same seeing (FWHM).  Whether this is caused by differences in the
instrument, guiding or atmosphere is currently not known.

The small systematic errors of UCAC based positions of randomly
selected Hipparcos stars confirms the good correction of UCAC3
epoch positions as function of magnitude and color, at least for
the 8 to 12 magnitude range. 
With UCAC3 positions agreeing with Hipparcos data the magnitude
equation seen in residuals with respect to Tycho-2 is an indication
for such small systematic errors in the Tycho-2 catalog.
These are likely introduced through the proper motions, thus the
early epoch, ground-based data, as also indicated in the
UCAC3 release paper.

The random errors in the observed UCAC$-$Hipparcos position
differences are even slightly smaller than the expected, combined
formal errors.  For the new Hipparcos reductions the difference is
only a few percent, while for the comparison with the original
Hipparcos Catalogue data the observed errors are about 10\% smaller
than expected.  This indicates a slightly overestimated error in the
original Hipparcos Catalogue proper motions.  The formal position
errors for the individual UCAC observations do include the formal
image profile fit error, and the conventional plate adjustment error
propagation.  The weighting scheme used in this individual CCD frame
least-square adjustments also include an estimated error contribution
from the turbulence in the atmosphere, scaled by the exposure time.
The mismatch between the actual PSF and the image profile model can
lead to an overestimation of the center position errors, particularly
for stars as bright as this sample of Hipparcos stars, which would
explain the slightly smaller than expected scatter in the Hipparcos to
UCAC position differences.  The exclusion of outliers at an arbitrary
limit of 200 mas could be another possible explanation. At the faint
end of Hipparcos (11th magnitude) the Hipparcos catalog positions are
of comparable precision to typical mean UCAC positions (based on 4
images) at their about 2000 epoch.  The next step after UCAC, the USNO
Robotic Astrometric Telescope (URAT) program \citep{urat} to begin in
2010 thus will likely be capable of improving proper motions of
individual Hipparcos stars significantly.


\acknowledgments

The entire UCAC team is thanked for making this all-sky survey a reality.
For more detailed information about ``who is who" in the UCAC project
the reader is referred to the readme file and UCAC3 release paper.
The California Institute of Technology is acknowledged for the 
{\em pgplot} software.
More information about this project is available at \\
\url{http://www.usno.navy.mil/usno/astrometry/}.


\clearpage


\clearpage



\begin{deluxetable}{crrrr}
\tabletypesize{\scriptsize}
\tablecaption{Summary of UCAC frames \tablenotemark{a}\label{framest1}} 
\tablewidth{0pt}

\tablehead{
\colhead{Site/orientation}       &
\colhead{number of}              &
\colhead{number of}              &
\colhead{number of}              &
\colhead{number of}             \\

\colhead{}                       &
\colhead{Calibration Frames}     &
\colhead{Survey Frames}          &
\colhead{Minor Planet Frames}    &
\colhead{Pluto Frames}           }

\startdata

CTIO east & 1582 &      5 &   14 &  0    \\
          &   0  &      3 &   14 &  0    \\
CTIO west & 1583 & 163460 &  828 & 10    \\
          &   0  & 155143 &  796 &  0    \\ 
NOFS east & 2452 &  66940 & 1340 & 84    \\
          &   0  &  58523 & 1156 & 74    \\ 
NOFS west & 1525 &   2580 &   32 &  0    \\
          &   0  &   2397 &   28 &  0    \\ 
\tableline
\tableline \\

Total     & 7142 & 232985 & 2214 & 94   \\ 
          &    0 & 216066 & 2068 & 74   \\ 
\enddata

\tablenotetext{a}{First row number represents the total number of UCAC
frames while the second row number gives the
number of frames used in the final UCAC3 reduction.}

\end{deluxetable}


\begin{deluxetable}{crr}
\tabletypesize{\scriptsize}
\tablecaption{Number of reference star observations used for 
    reductions \label{framest2}}

\tablewidth{0pt}

\tablehead{
\colhead{}                         &
\colhead{number of}                &
\colhead{number of}               \\
			           
\colhead{site/orientation}         &
\colhead{2MASS star}               &
\colhead{Tycho-2 star}            \\
			           
\colhead{}                         &
\colhead{observations}             &
\colhead{observations}             }

\startdata

CTIO east &   8772812  &    1600   \\
CTIO west & 203915168  & 9683015   \\
NOFS east &  58493251  & 4071859   \\
NOFS west &   7856501  &  186557   \\
\tableline
\tableline \\
Total     & 279037732  & 13943031  \\

\enddata
\end{deluxetable}


\begin{deluxetable}{lccccccccccccccccc}
\tabletypesize{\tiny}\tablecaption{Example CTIO west CTE lookup table
for frames taken at 20 second exposures with corrections given in
mas\label{CTEt1}}

\tablewidth{0pt}

\tablehead{
\colhead{}                 &
\colhead{ 8.0}             &
\colhead{ 8.5}             &
\colhead{ 9.0}             &
\colhead{ 9.5}             &
\colhead{10.0}             &
\colhead{10.5}             &
\colhead{11.0}             &
\colhead{11.5}             &
\colhead{12.0}             &
\colhead{12.5}             &
\colhead{13.0}             &
\colhead{13.5}             &
\colhead{14.0}             &
\colhead{14.5}             &
\colhead{15.0}             &
\colhead{15.5}             &
\colhead{16.0}            \\
	                  
\colhead{}                 &
\colhead{mag}              &
\colhead{mag}              &
\colhead{mag}              &
\colhead{mag}              &
\colhead{mag}              &
\colhead{mag}              &
\colhead{mag}              &
\colhead{mag}              &
\colhead{mag}              &
\colhead{mag}              &
\colhead{mag}              &
\colhead{mag}              &
\colhead{mag}              &
\colhead{mag}              &
\colhead{mag}              &
\colhead{mag}              &
\colhead{mag}              }

\startdata

$x bin1$ & 147 & 148 & 143 & 134 & 119 & 108 & 95 & 81 & 66 & 49 & 33 & 18 & -8 & -44 & -69 & -76 & -86  \\
$x bin2$ & 116 & 111 & 115 & 111 & 105 &  90 & 80 & 69 & 53 & 42 & 27 & 18 & -4 & -32 & -52 & -59 & -74  \\
$x bin3$ & 101 &  92 &  95 &  85 &  76 &  67 & 60 & 44 & 41 & 33 & 24 & 14 &  2 & -17 & -31 & -39 & -48  \\
$x bin4$ &  68 &  61 &  61 &  59 &  56 &  43 & 38 & 33 & 29 & 23 & 19 & 11 &  2 & -10 & -17 & -21 & -30  \\
$x bin5$ &  52 &  47 &  42 &  41 &  43 &  37 & 32 & 28 & 25 & 23 & 17 & 13 & 11 &   3 &   1 &   0 & -10  \\

\enddata                             
\end{deluxetable}


\begin{table}
\begin{center}
\caption{RMS position differences and expected errors of observed
         UCAC3 $-$ Hipparcos positions.\label{TYCt1}}
\vspace*{5mm}
\begin{tabular}{rcccccc}
\tableline\tableline
       &            &  difference   & Hipparcos err &   UCAC3 err   & combined err  &  ratio      \\
number & Vmag range &  RA     Dec   &  RA      Dec  &  RA      Dec  &  RA     Dec   & RA   Dec    \\
observ.&    mag     &  mas    mas   &  mas     mas  &  mas     mas  &  mas    mas   & diff/c.err  \\
\tableline
\multicolumn{7}{c}{Hipparcos (original)}\\
\tableline
 423&\phn8.0 $-$\phn8.5& 41.9\phn46.3& 10.7\phn\phn8.8& 46.9\phn47.1& 48.1\phn47.9& 0.87\phn0.97\\
 717&\phn8.5 $-$\phn9.0& 45.2\phn41.9& 12.7\phn   11.7& 47.8\phn48.0& 49.4\phn49.4& 0.92\phn0.85\\
 802&\phn9.0 $-$   10.0& 43.6\phn44.6& 21.8\phn   18.3& 47.3\phn47.6& 52.1\phn51.0& 0.84\phn0.88\\
1227&   10.0 $-$   11.0& 45.6\phn45.5& 28.3\phn   22.5& 43.3\phn43.9& 51.7\phn49.3& 0.88\phn0.92\\
 978&   11.0 $-$   99.0& 52.8\phn48.2& 42.7\phn   35.7& 42.2\phn42.7& 60.0\phn55.6& 0.88\phn0.87\\
\tableline
\multicolumn{7}{c}{Hipparcos (new reduction)}\\
\tableline
 423&\phn8.0 $-$\phn8.5& 42.3\phn45.8&\phn 8.9\phn\phn7.3& 46.9\phn47.1& 47.7\phn47.6& 0.89\phn0.96\\
 722&\phn8.5 $-$\phn9.0& 46.4\phn44.2&    12.0\phn   10.0& 47.8\phn48.0& 49.3\phn49.1& 0.94\phn0.90\\
 801&\phn9.0 $-$   10.0& 45.3\phn42.6&    14.0\phn   11.6& 47.3\phn47.5& 49.3\phn48.9& 0.92\phn0.87\\
1238&   10.0 $-$   11.0& 47.4\phn44.1&    20.8\phn   16.9& 43.4\phn43.9& 48.1\phn47.1& 0.99\phn0.94\\
 993&   11.0 $-$   99.0& 52.3\phn49.9&    36.7\phn   27.5& 42.2\phn42.8& 56.0\phn50.9& 0.93\phn0.98\\
\tableline

\end{tabular}
\end{center}
\end{table}


\end{document}